\documentclass[11pt,fleqn]{article}
\usepackage{epsfig}
\newcommand{\la}{\label}
\newcommand{\re}{\ref}
\newcommand{\ci}{\cite}
\newcommand{\beqn}{\begin{eqnarray}}
\newcommand{\eeqn}{\end{eqnarray}}
\newcommand{\bequ}{\begin{equation}}
\newcommand{\eequ}{\end{equation}}
\newcommand{\bsl}{\begin{sloppypar}}
\newcommand{\esl}{\end{sloppypar}}

\begin{document}
\null
\hfill DESY 00-002\\
\vspace{.1cm}\hfill UWThPh-2000-3\\
\vspace{.1cm}\hfill WUE-ITP-2000-004\\
\vspace{.1cm}\hfill HEPHY-PUB 726/2000\\
\vspace{.1cm}\hfill hep-ph/0002253\\
\vskip .4cm

\begin{center}
{\Large \bf Exploiting Spin Correlations in\\[.4em]
Neutralino Production 
and Decay\\[.4em]
with Polarized $e^-$ and $e^+$ Beams 
\footnote{Contribution to the Proceedings of the 2nd Joint ECFA/DESY study on
Physics\\\mbox{\hspace{.5cm}} and Detectors for a Linear Electron-Positron Collider}%
}
\end{center}
\vskip 1.5em
{\large
{\sc G.~Moortgat--Pick$^{a}$\footnote{e-mail:
    gudrid@mail.desy.de},
A.~Bartl$^{b}$\footnote{e-mail:
     bartl@ap.univie.ac.at},  
H. Fraas$^{c}$\footnote{e-mail:
    fraas@physik.uni-wuerzburg.de},
W.~Majerotto$^{d}$\footnote{e-mail:
     majer@qhepu3.oeaw.ac.at.}%
}}\\[3ex]
{\footnotesize \it
$^{a}$ DESY, Deutsches Elektronen--Synchrotron, D--22603 Germany}\\
{\footnotesize \it
$^{b}$ Institut f\"ur Theoretische Physik, Universit\"at Wien, 
A--1090 Wien, Austria}\\
{\footnotesize \it
$^{c}$ Institut f\"ur Theoretische Physik, Universit\"at
W\"urzburg, D--97074 W\"urzburg, Germany}\\
{\footnotesize \it
$^{d}$ Institut f\"ur Hochenergiephysik der \"Osterreichischen
 Akademie der Wissenschaften,\\\phantom{$^{d}$} A--1050 Wien, Austria}
\vskip .5em
\par
\vskip .4cm

\begin{abstract}
We study the production process $e^+ e^-\to \tilde{\chi}^0_1 \tilde{\chi}^0_2$
and the subsequent decay $\tilde{\chi}^0_2\to \tilde{\chi}^0_1 \ell^+ \ell^-$ 
with polarized $e^+$ and $e^-$ beams, including the spin correlations 
between production and decay. We work out the advantages of 
polarizing both beams. 
We study in detail the angular distribution and the forward--backward 
asymmetry of the decay lepton as well as the opening angle distribution 
between the decay leptons. 
We investigate the dependence on 
the masses of $\tilde{e}_L$ and $\tilde{e}_R$ and on the mixing 
character of the neutralinos. In particular we study the dependence
on the gaugino mass parameter $M_1$.
\end{abstract}

\vspace{1em}
\hfill

\section{Introduction}
The search for supersymmetric (SUSY) particles and the determination of their
properties will be one of the main goals of a future $e^+e^-$ linear collider.
Particularly interesting will be the experimental study of the neutralinos, 
which are the quantum mechanical mixtures of the neutral gauginos and
higgsinos, the SUSY partners of the neutral gauge and Higgs bosons. In the
Minimal Supersymmetric Standard Model (MSSM) there are four neutralinos
$\tilde{\chi}_i^0$, $i =1,\ldots,4$. 
In extensions of the MSSM there may be more
than four neutralinos. Usually the lightest neutralino $\tilde{\chi}_1^0$ is
the lightest SUSY particle LSP. Therefore, the production
of the lightest and the second lightest neutralino
can presumably be studied at an $e^+e^-$ linear collider with
a CMS energy $\sqrt{s} = 500$~GeV. 
The aim will be to precisely determine
the SUSY parameters of the neutralino system. 
By a detailed study of the 
neutralino system one can also examine the question whether the MSSM or 
another SUSY model is realized in nature. Models with an extended neutralino 
sector have been discussed in \ci{Gudi_sitges}.
Recently a method for determining the SUSY parameter $M_2$, $\mu$ and 
$\tan\beta$ by measuring suitable observables in chargino production
$e^+ e^-\to \tilde{\chi}^+_i \tilde{\chi}^-_j$ ($i,j=1,2$) has been proposed
in \ci{Choi}. 
The process $e^+ e^-\to \tilde{\chi}^+_i \tilde{\chi}^-_j$, 
$\tilde{\chi}^+_i\to \tilde{\chi}^0_k \ell^+ \nu$ including the full spin 
correlation has been studied  in \ci{Gudi_char}.

In previous papers (see \ci{Hans_nucl,Ambrosanio} and references therein) 
the cross sections of $e^+e^- \to \tilde{\chi}_i^0 
\tilde{\chi}_j^0$, $i,j = 1 \ldots 4$, and the branching ratios and energy 
distributions of neutralino decays were studied, which do not depend on
spin correlations. In the calculation of the decay angular distributions,
however, one has to take into account the spin correlations between
production and decay of the neutralinos. 
In \ci{Gudi_physrev} 
we have studied the
process $e^+ e^- \to \tilde{\chi}_i^0 \tilde{\chi}_j^0$, 
$i,j = 1 \ldots 4$, with
unpolarized beams and the subsequent leptonic decays $\tilde{\chi}_i^0 \to
\tilde{\chi}_k^0 \ell^+ \ell^-$ including the complete spin correlations. In
\ci{Gudi_neut} 
we have given the complete analytical formulae for longitudinally
polarized beams,
fully including the spin correlations between production and decay. The
formulae have been given in the laboratory system in terms of the basic
kinematic variables. 

In the present paper we extend our analysis in \ci{Gudi_neut} 
and study neutralino
production in the case that both beams are polarized. We
show that by a suitable choice of the $e^+$ beam polarization not only higher
cross sections but also additional information can be gained. We give numerical
predictions for the process $e^+ e^- \to \tilde{\chi}_1^0 \tilde{\chi}_2^0$ 
with
$\tilde{\chi}_2^0 \to \tilde{\chi}_1^0 e^+ e^-$ including the full spin
correlations between production and decay. The framework of our studies is
the MSSM. We first assume the GUT relation $M_1/M_2 = \frac{5}{3} \tan^2
\Theta_W$ for the gaugino mass parameters (note that in Refs. 
\ci{Hans_nucl, Gudi_physrev, Gudi_neut} we
used the notation $M'$ and $M$ for $M_1$ and $M_2$). We consider two
gaugino--like and one higgsino--like scenario and study the dependence 
of the cross section, the forward--backward asymmetry of the decay electron, 
and the opening angle distribution of the decay lepton pair $e^+$ and $e^-$ 
on the beam polarizations and on the 
masses of the exchanged $\tilde{e}_L$ and $\tilde{e}_R$. Then we relax the GUT 
relation between $M_1$ and $M_2$ and study the $M_1$ dependence of
$\sigma(e^+ e^- \to \tilde{\chi}_1^0 \tilde{\chi}_2^0) \times 
BR (\tilde{\chi}_2^0 \to \tilde{\chi}_1^0 e^+ e^-)$ and the 
forward-backward asymmetry of the decay electron for
various beam polarizations and slepton masses. We also discuss which 
observables are suitable for the determination of the neutralino
parameters and selectron masses.
\section{Spin correlations between production and decay}\la{sec:1}
\mathindent0cm
\bsl
Both the helicity amplitudes $T_P^{\lambda_i\lambda_j}$ for the 
production process 
$e^{-}(p_1)e^{+}(p_2)\to \tilde{\chi}^0_i(p_3,\lambda_i)
\tilde{\chi}^0_j(p_4,\lambda_j)$ and the 
helicity amplitudes $T_{D,\lambda_i}$ and $T_{D,\lambda_j}$ for the decay 
processes $\tilde{\chi}^0_i(p_3,\lambda_i)\to\tilde{\chi}^0_k(p_5) \ell^+(p_6) 
\ell^{-}(p_7)$ 
and $\tilde{\chi}^0_j(p_4,\lambda_j)\to\tilde{\chi}^0_l(p_8) \ell^+(p_9) 
\ell^{-}(p_{10})$,
respectively, receive contributions from $Z^0$ exchange in the direct channel 
and from $\tilde{e}_{L,R}$ exchange in the crossed channels. 
\esl

The amplitude 
squared of the combined process of production and decay is:
\bequ
|T|^2=|\Delta(\tilde{\chi}_i^0)|^2 |\Delta(\tilde{\chi}_j^0)|^2 
\rho_P^{\lambda_i \lambda_j \lambda^{'}_i \lambda^{'}_j} 
\rho_{D, \lambda^{'}_i\lambda_i}\rho_{D,\lambda_j^{'} \lambda_j}
\hspace{.3cm}\mbox{(summed over helicities)}.
\la{eq4_4e}
\eequ
It is composed of the (unnormalized) spin density production matrix
\bequ
\rho_P^{\lambda_i\lambda_j\lambda_i^{'}\lambda_j^{'}}=T_P^{\lambda_i\lambda_j}
T_P^{\lambda_i^{'}\lambda_j^{'}*},
\la{eq4_4f}
\eequ
the decay matrices
\bequ
\rho_{D, \lambda_i^{'} \lambda_i}=T_{D, \lambda_i} T_{D, \lambda_i^{'}}^{*}
\quad\mbox{and}\quad
\rho_{D, \lambda_j^{'} \lambda_j}=T_{D,\lambda_j} T_{D,\lambda_j^{'}}^*,
\la{eq4_4g}
\eequ
and the propagators 
\bequ
\Delta(\tilde{\chi}^0_{i,j})=1/[p^2_{3,4}-m_{i,j}^2+i m_{i,j} \Gamma_{i,j,}].
\eequ
Here $p_{3,4}^2$, $\lambda_{i,j}$, $m_{i,j}$ and $\Gamma_{i,j}$ denote the 
four--momentum squared, helicity, mass and total width of 
$\tilde{\chi}^0_{i,j}$. For these propagators we use the narrow--width 
approximation. 

\bsl
Introducing a suitable set of polarization vectors for each of the 
neutralinos the density matrices can be expanded in terms of Pauli matrices 
\nolinebreak$\sigma^a$: \esl
\beqn 
\rho_P^{\lambda_i\lambda_j\lambda_i^{'}\lambda_j^{'}}&=&
(\delta_{\lambda_i\lambda_i^{'}} \delta_{\lambda_j\lambda_j^{'}}
P(\tilde{\chi}^0_i \tilde{\chi}^0_j)
+\delta_{\lambda_j\lambda_j^{'}}\sum_a
\sigma^a_{\lambda_i\lambda_i^{'}}\Sigma^a_P(\tilde{\chi}^0_i)
\nonumber\\ &&
+\delta_{\lambda_i\lambda_i^{'}}\sum_b
\sigma^b_{\lambda_j\lambda_j^{'}}\Sigma^b_P(\tilde{\chi}^0_j)
+\sum_{ab}\sigma^a_{\lambda_i\lambda_i^{'}}
\sigma^b_{\lambda_j\lambda_j^{'}}
\Sigma^{ab}_P(\tilde{\chi}^0_i \tilde{\chi}^0_j)),\la{eq4_4h}\\
\rho_{D,\lambda_i^{'}\lambda_i}&=&(\delta_{\lambda_i^{'}\lambda_i}
D(\tilde{\chi}^0_i)+\sum_a
\sigma^a_{\lambda_i^{'}\lambda_i} \Sigma^a_D(\tilde{\chi}^0_i)),\la{eq4_4i}\\
\rho_{D,\lambda_j^{'}\lambda_j}&=&(\delta_{\lambda_j^{'}\lambda_j}
D(\tilde{\chi}^0_j)+\sum_b
\sigma^b_{\lambda_j^{'}\lambda_j} \Sigma^b_D(\tilde{\chi}^0_j)) \quad 
\mbox{, a, b=1,2,3}.\la{eq4_4j}
\eeqn
We choose the polarization vectors such that $\Sigma^1_P(\tilde{\chi}^0_{i,j})$
 describes the 
transverse polarization in the production plane, 
$\Sigma^2_P(\tilde{\chi}^0_{i,j})$ denotes the 
polarization perpendicular to the production plane and 
$\Sigma^3_P(\tilde{\chi}^0_{i,j})$ describes 
the longitudinal polarization of the respective neutralino. 
$\Sigma^{ab}_P(\tilde{\chi}^0_i\tilde{\chi}^0_j)$ 
is due to correlations between 
the polarizations of both neutralinos. The complete analytical expressions for 
the production density matrix and for the decay matrices are given in
\ci{Gudi_neut}.

If CP is conserved the neutralino couplings are real. It can be seen in 
\ci{Gudi_neut} that in this case, due to the 
Majorana character of the neutralinos, the quantities $P$, $\Sigma^1_P$, 
$\Sigma^{11}_P$, $\Sigma^{22}_P$, $\Sigma^{33}_P$, $\Sigma^{23}_P$ are 
forward--backward symmetric, whereas
the quantities $\Sigma^{2}_P$, $\Sigma^{3}_P$, $\Sigma^{12}_P$, $\Sigma^{13}_P$
are forward--backward antisymmetric.

The amplitude squared of the combined process of production and decay can
be written as:
\beqn
&&\mbox{\hspace{-.7cm}}
|T|^2=4|\Delta(\tilde{\chi}^0_i)|^2|
\Delta(\tilde{\chi}^0_j)|^2
         \Big(P(\tilde{\chi}^0_i \tilde{\chi}^0_j) 
D(\tilde{\chi}^0_i) D(\tilde{\chi}^0_j)
    +\sum^3_{a=1}\Sigma_P^a(\tilde{\chi}^0_i) 
\Sigma_D^a(\tilde{\chi}^0_i) 
D(\tilde{\chi}^0_j)\nonumber\\
& &\phantom{4|}
+\sum^3_{b=1}\Sigma_P^b(\tilde{\chi}^0_j) \Sigma_D^b(\tilde{\chi}^0_j)
D(\tilde{\chi}^0_i)
    +\sum^3_{a,b=1}\Sigma_P^{ab}(\tilde{\chi}^0_i\tilde{\chi}^0_j)
 \Sigma^a_D(\tilde{\chi}^0_i) 
\Sigma^b_D(\tilde{\chi}^0_j)\Big).
\la{eq4_5}
\eeqn
The differential cross section in the laboratory system 
is then given by
\begin{equation}
d\sigma_e=\frac{1}{2 s}|T|^2 (2\pi)^4
\delta^4(p_1+p_2-\sum_{i} p_i) d{\rm lips}(p_3\ldots p_{10})\label{eq_13},
\end{equation}
$d{\rm lips}(p_3,\ldots,p_{10})$ 
is the Lorentz invariant phase space element.

If one neglects all spin correlations between production and decay only the 
first term in (\re{eq4_5}) contributes. The second and third term in 
(\re{eq4_5}) describe the spin correlations between the production and the 
decay process  and the last term is due to spin--spin correlations between 
both decaying neutralinos.
\section{Numerical analysis and discussion}\la{sec:3}
In the MSSM \ci{Haber-Kane} 
the masses and couplings of neutralinos are determined
 by the 
parameters $M_1$, $M_2$, $\mu$, $\tan\beta$, which can be chosen real if CP 
violation is neglected. Moreover, one usually makes use of the GUT relation 
\bequ
M_1=\frac{5}{3} M_2 \tan^2\Theta_W.
\la{eq_m1}
\eequ
The neutralino mass mixing matrix in 
the convention used can be found in \ci{Hans_nucl}.

The total cross section of $e^+e^-\to\tilde{\chi}^0_i\tilde{\chi}^0_j$ and 
the decay rate $\tilde{\chi}^0_i\to\tilde{\chi}^0_k e^+ e^-$ further depend 
on the masses of $\tilde{e}_L$ and $\tilde{e}_R$.
In the following numerical analysis we study neutralino production and decay 
in three scenarios, which we denote by A1, A2, and B. 
The corresponding parameters are given in 
Table~\ref{tab:1}.

Scenario~A1 corresponds to that in \ci{Blair}.
In this scenario $\tilde{\chi}^0_1$ is $\tilde{B}$--like and $\tilde{\chi}^0_2$
 is $\tilde{W}^3$--like. The selectron masses are 
$m_{\tilde{e}_L}=176$~GeV, $m_{\tilde{e}_R}=132$~GeV.
Scenario~A2 differs from A1 only by the mass of $\tilde{e}_L$. 
In scenario~B the same masses as in scenario~A1 are taken for 
$\tilde{\chi}^0_1$, $\tilde{\chi}^0_2$, $\tilde{e}_L$, $\tilde{e}_R$.
$\tilde{\chi}^0_1$ and $\tilde{\chi}^0_2$ are, however, higgsino--like
due to the choice $\mu<M_2$.
\subsection{Effects of  beam polarizations on the total cross section}
\la{sec:31}
The cross section $\sigma(e^+ e^-\to \tilde{\chi}^0_1 \tilde{\chi}^0_2)$ is 
shown in Figs.~1a, 1b, and 1c as 
a function of the longitudinal beam polarizations 
$P_-^3$ for electrons and $P_+^3$ for positrons, for scenario~A1, A2 and B, 
respectively, (with $P_{\pm}^3=\{-1,0,1\}$ for 
$\{$left--, un--, right--$\}$polarized ). The cross section 
$\sigma(e^+e^-\to \tilde{\chi}^0_1 \tilde{\chi}^0_2)$
is shown at 
$\sqrt{s}=(m_{\tilde{\chi}^0_1}+m_{\tilde{\chi}^0_2})+30$~GeV.


In Figs.~1a, 1b, and 1c the white area is covered by an electron 
polarization $|P_-^3|\le 85\%$ and a positron polarization $|P_+^3|\le 60\%$.
The cross section can be enhanced by a factor 2--3 by polarizing both beams. 
Theoretically, for pure gaugino--like neutralinos and
$m_{\tilde{e}_L}\gg m_{\tilde{e}_R}$ ($m_{\tilde{e}_L}\ll m_{\tilde{e}_R}$)
and  $P_-^3=+1$, $P_+^3=-1$ ($P_-^3=-1$, $P_+^3=+1$), the cross section
could be enlarged by a factor 4. For pure higgsino--like neutralinos
and $P_-^3=+1$, $P_+^3=-1$ ($P_-^3=-1$, $P_+^3=+1$) this factor would be 
1.7 (2.3) \ci{Gudi_diss}.

One clearly recognizes, Figs.~1a, 1b, and 1c,  
the sensitive dependence of the cross section
on the selectron masses 
$m_{\tilde{e}_L}$ and $m_{\tilde{e}_R}$ as well as on the mixing character of 
$\tilde{\chi}^0_1$ and $\tilde{\chi}^0_2$. In scenario~A2 with 
$m_{\tilde{e}_L} \gg m_{\tilde{e}_R}$ and gaugino--like 
$\tilde{\chi}^0_{1,2}$, one expects the largest cross section for $P_-^3=+1$ 
and $P_+^3=-1$, see Fig.~1b. 
In scenario~B the cross  section is governed by $Z^0$ exchange and
is therefore rather symmetric for $P_-^3=\pm \leftrightarrow P_+^3=\mp$.
The beam polarizations are 
a useful tool for getting more 
information about $m_{\tilde{e}_L}$ and $m_{\tilde{e}_R}$.

If the polarizations of both beams are varied, the relative size of the cross 
sections strongly depends on the mixing character of both
neutralinos $\tilde{\chi}^0_{1,2}$ and on the selectron masses
\ci{Gudi_diss}. If $\tilde{\chi}^0_1$ and $\tilde{\chi}^0_2$ are pure 
higgsinos, one obtains for $|P_-^3|=85\%$ and $|P_+^3|=60\%$ the sequence 
\bequ
\sigma_e^{-+}>\sigma_e^{+-}>\sigma_e^{-0}>
\sigma_e^{00}>\sigma_e^{+0}>\sigma_e^{--}>\sigma_e^{++}.
\la{eq_pol1}
\eequ
Here $\sigma_e=\sigma(e^-e^+\to \tilde{\chi}^0_1 \tilde{\chi}^0_2)\times
BR(\tilde{\chi}^0_2\to \tilde{\chi}^0_1 \ell^+ \ell^-)$ and
$(-+)$ etc. denotes the sign of the electron polarization $P_-^3$ and of 
the positron polarization $P_+^3$, respectively.

If $\tilde{\chi}^0_1$ and $\tilde{\chi}^0_2$ are 
pure gauginos the order of the cross sections depends on the relative 
magnitude of the selectron masses $m_{\tilde{e}_L}$ and 
$m_{\tilde{e}_R}$.
For $m_{\tilde{e}_L}\gg m_{\tilde{e}_R}$ only right selectron exchange 
contributes, and one obtains
\bequ
\sigma_e^{+-}>\sigma_e^{+0}>\sigma_e^{00}>\sigma_e^{++}>\sigma_e^{--}>
\sigma_e^{-0}>\sigma_e^{-+},
\la{eq_pol3}
\eequ
whereas for $m_{\tilde{e}_R}\gg m_{\tilde{e}_L}$, one gets:  
\bequ
\sigma_e^{-+}>\sigma_e^{-0}>\sigma_e^{00}>\sigma_e^{--}>\sigma_e^{++}>
\sigma_e^{+0}>\sigma_e^{+-}.
\la{eq_pol2}
\eequ
The case of a heavy right slepton may be realized in extended SUSY models
(\ci{Hesselbach} and references therein).

Comparing (\re{eq_pol1}) and (\re{eq_pol2}) shows that polarizing both beams
allows one to distinguish between a higgsino--like scenario and
a gaugino--like scenario with dominating $\tilde{e}_L$ exchange. 
This is not possible if only the electron beam is polarized. 

In Table~\re{table_2} we show the cross sections for various polarization
configurations for our scenarios A1, A2, and B 
at $\sqrt{s}=(m_{\tilde{\chi}^0_1}+m_{\tilde{\chi}^0_2})+30$~GeV 
for $P_{-}^3=0, \pm 85\%$ and $P_+^3=0, \pm 60\%$.
For the higgsino--like scenario~B the sequence of the cross sections
coincides with (\re{eq_pol1}).
One notices that one obtains the same ordering of the polarized cross
sections for the gaugino--like scenario~A1.
This shows that the relative size of the cross sections sensitively depends 
on the mass difference between $\tilde{e}_L$ and $\tilde{e}_R$, 
which is rather small in 
scenario~A1, $m_{\tilde{e}_L}-m_{\tilde{e}_R}=44$~GeV. Comparing the 
sequence of cross sections 
for the gaugino--like scenario~A2, see Table~\re{table_2}, 
with that for pure 
$\tilde{e}_R$ exchange, see (\re{eq_pol3}), one sees a small influence of 
$\tilde{e}_L$ exchange despite the rather high $\tilde{e}_L$ mass,
$m_{\tilde{e}_L}=500$~GeV.
\subsection{Lepton forward--backward asymmetry}\la{sec:32}
Owing to the Majorana character of the neutralinos the angular distribution 
of the production process 
is forward--backward symmetric \ci{Christova}.
The angular distribution of the decay lepton, however, depends sensitively on 
the polarization of $\tilde{\chi}^0_i$.
Since the longitudinal polarization $\Sigma^3_P$ and the transverse 
polarization $\Sigma^1_P$ of $\tilde{\chi}^0_i$ are forward--backward 
antisymmetric, the lepton forward--backward asymmetry $A_{FB}$
of the decay lepton may become quite large.
The lepton forward--backward asymmetry $A_{FB}$ is defined as 
\bequ
A_{FB}=\frac{\sigma_e(\cos\Theta_e >0)-\sigma_e(\cos\Theta_e<0)}
{\sigma_e(\cos\Theta_e >0)+\sigma_e(\cos\Theta_e<0)},
\la{eq_30}
\eequ
where $\sigma_e$ is a short--hand notation for
$\sigma_e=\sigma(e^+ e^-\to \tilde{\chi}^0_1 \tilde{\chi}^0_2)\times
BR(\tilde{\chi}^0_2 \to \tilde{\chi}^0_1 e^+ e^-)$.
We will show $A_{FB}$ not too far from threshold because  it 
decreases with $\sqrt{s}$ for fixed neutralino masses.

In Figs.~2a, 2b, and 2c we show $A_{FB}$ of 
the decay electron as a function of the electron and positron polarizations
 for the scenarios~A1, A2, and B, respectively, at 
$\sqrt{s}=(m_{\tilde{\chi}^0_1}+m_{\tilde{\chi}^0_2})+30$~GeV. First
one notices that polarizing suitably both beams
gives a larger asymmetry. Actually,
in the scenarios A1 and B 
$A_{FB}$ turns out to be practically zero for both beams unpolarized.
The figures also exhibit a very different pattern. When comparing 
Fig.~2a with Fig.~2b, the different behaviour
is due to the different masses of $\tilde{e}_L$.
In the higgsino scenario~B the asymmetries are much smaller. Notice
again the symmetry of $P_{-}^3=\pm\leftrightarrow P_+^3=\mp$ in scenario~B as 
already observed in the total cross section.
Measuring the lepton forward--backward asymmetry $A_{FB}$ in addition to the 
total cross section strongly constrains the selectron masses $m_{\tilde{e}_L}$
and $m_{\tilde{e}_R}$ as well as the mixing properties of
$\tilde{\chi}^0_1$ and $\tilde{\chi}^0_2$.

We also studied the dependence of the lepton forward--backward asymmetry 
$A_{FB}$ on $\sqrt{s}$. For 
$\sqrt{s}\gg (m_{\tilde{\chi}^0_1}+m_{\tilde{\chi}^0_2})$ the
angular distribution of the decay lepton
is essentially the same as that of the decaying 
neutralino $\tilde{\chi}^0_2$ \ci{Feng}.
Therefore the lepton forward--backward asymmetry practically vanishes.
\begin{table}
\begin{tabular}{|l||c|c|c|c|c|c|c|c|c|c|}
 & $M_2$ & $\mu$ & $m_{\tilde{e}_L}$ & $m_{\tilde{e}_R}$ & 
$m_{\tilde{\chi}^{0}_1}$ &
$m_{\tilde{\chi}^{0}_2}$ & $\Gamma_{\tilde{\chi}^{0}_2}$ & 
$O^{''L}_{12}$ & $f^{L}_{\ell 1}f^{L}_{\ell 2}$ &
$f^{R}_{\ell 1}f^{R}_{\ell 2}$\\ \hline
 A1 & 152 & 316 & 176 & 132 & 71 & 130 & 25E$-6$ & $-.02$ & $-.20$ 
& $-.12$\\ \hline
 A2 & 152 & 316 & 500 & 132 & 71 & 130 & 15E$-6$  & $-.02$ & $-.20$ 
& $-.12$\\ \hline
B & 250 & 125 & 176 & 132 & 71 & 130 & 369E$-6$ & +.39 &  2E$-5$ & .027\\ 
\hline
\end{tabular}
\caption{Parameters, masses, and total $\tilde{\chi}^0_2$ 
width (in GeV) and couplings
in scenarios A1, A2, and B for $\tan\beta=3$. \label{tab:1}}
\end{table}

\begin{table}
\begin{center}
\begin{tabular}{|l||c|c|c|c|c|c|c|}
 & \multicolumn{7}{|c|}
{$\sqrt{s}=(m_{\tilde{\chi}^0_1}+m_{\tilde{\chi}^0_2})+30$~GeV} \\ \hline
A1 & $(-+)$ & $(+-)$ &$(-0)$ &$(00)$ & $(+0)$ &$(--)$ & $(++)$\\ 
\hline
 $\sigma_e$/fb 
& 10.2 & 6.7 & 6.6 & 5.6 & 4.6 & 3.1 & 2.5\\ \hline
A2 & $(+-)$ & $(+0)$ &$(00)$ &$(++)$ & $(-+)$ & $(-0)$ & $(--)$\\ 
\hline
 $\sigma_e$/fb 
 & 9.5 & 6.0 & 3.6 & 2.5 &1.3 &1.1 & 1.0\\ \hline
B & $(-+)$ & $(+-)$ &$(-0)$ &$(00)$ & $(+0)$ &$(--)$ & $(++)$\\ 
\hline
 $\sigma_e$/fb 
& 21.7 & 19.0 & 14.3 & 13.5 & 12.7 & 6.8 & 6.4\\ \hline
\end{tabular}
\caption{Polarized cross sections 
$\sigma_e=\sigma(e^+ e^-\to \tilde{\chi}^0_1 \tilde{\chi}^0_2)
\times BR(\tilde{\chi}^0_2\to \tilde{\chi}^0_1 e^+ e^-)$/fb
at $\sqrt{s}=m_{\tilde{\chi}^0_2}+m_{\tilde{\chi}^0_1}+30$~GeV
in scenarios~A1, A2, and B, see Table~\re{tab:1}, for unpolarized beams 
$(00)$, only 
electron beam polarized $(-0)$, $(+0)$ with  $P_1^3=\pm 85\%$
and both beams polarized with $P_1^3=-85\%$, $P_2^3=+60\%$ $(-+)$ and 
$P_1^3=+85\%$, $P_2^3=-60\%$ $(+-)$. \la{table_2}}
\end{center}
\end{table}
\subsection{Opening angle distribution}\la{sec:33}
\bsl
The opening angle distribution between the two 
leptons from the decay of one of 
the neutralinos, $\tilde{\chi}^0_2\to \ell^+ \ell^- \tilde{\chi}^0_1$,
is independent of the spin correlations 
due to the Majorana nature of the neutralinos \ci{Gudi_fac}.
Therefore, it factorizes into the 
contributions from production and decay. For the same reason this is also valid
for the energy distribution of the neutralino decay pro\-ducts. 
For both distributions it is 
suitable to parametrize the phase space by the scattering angle $\Theta$ 
between the incoming $e^{-}(p_1)$ beam and the outgoing neutralino 
$\tilde{\chi}^0_2(p_4)$, the azimuthal angle 
$\Phi_{\tilde{\chi}^0_2 \ell^{-}}$ 
between the scattering plane and the ($\tilde{\chi}^0_2 \ell^-$)--plane 
and the opening angle $\Theta_{+-}$ 
between the leptons $\ell^+$ and $\ell^-$ 
from the decay of the neutralino $\tilde{\chi}^0_2$.
Since the phase space is independent of the azimuthal angle, the contributions
of the transverse polarizations of the neutralino 
vanish after integration over $\Phi_{\tilde{\chi}^0_2 \ell^{-}}$.
As for the longitudinal polarization 
$\Sigma^3_P$, the Majorana character of 
the neutralino is crucial. If CP is conserved 
$\Sigma^3_P$ is forward--backward antisymmetric,
$\Sigma^3_P(-\cos\Theta)=-\Sigma^3_P(\cos\Theta)$, so that the contribution of 
the longitudinal polarization vanishes after integration over the 
scattering angle $\Theta$ \ci{Gudi_fac}. 
\esl

In Figs.~\re{fig_5}a and \re{fig_5}b we 
show the distribution of the angle $\Theta_{+-}$
between the decay leptons 
from $e^+e^-\to \tilde{\chi}^0_1 \tilde{\chi}^0_2$,
$\tilde{\chi}^0_2\to e^+ e^- \tilde{\chi}^0_1$ in the scenarios~A1, A2 and B at
$\sqrt{s}=(m_{\tilde{\chi}^0_1}+m_{\tilde{\chi}^0_2})+30$~GeV and for various
beam polarizations. 
One notices that the shape of the $\Theta_{+-}$ distribution 
is mainly determined by the mixing character of the neutralinos. The 
selectron masses mainly influence the size of the cross section.
Due to the factorization of production and decay
the beam polarizations have no influence on the 
shape of this distribution.

\begin{figure}[t]
\hspace{-.8cm}
\begin{minipage}{7cm}
\begin{picture}(7,7)
\put(0,0){\includegraphics{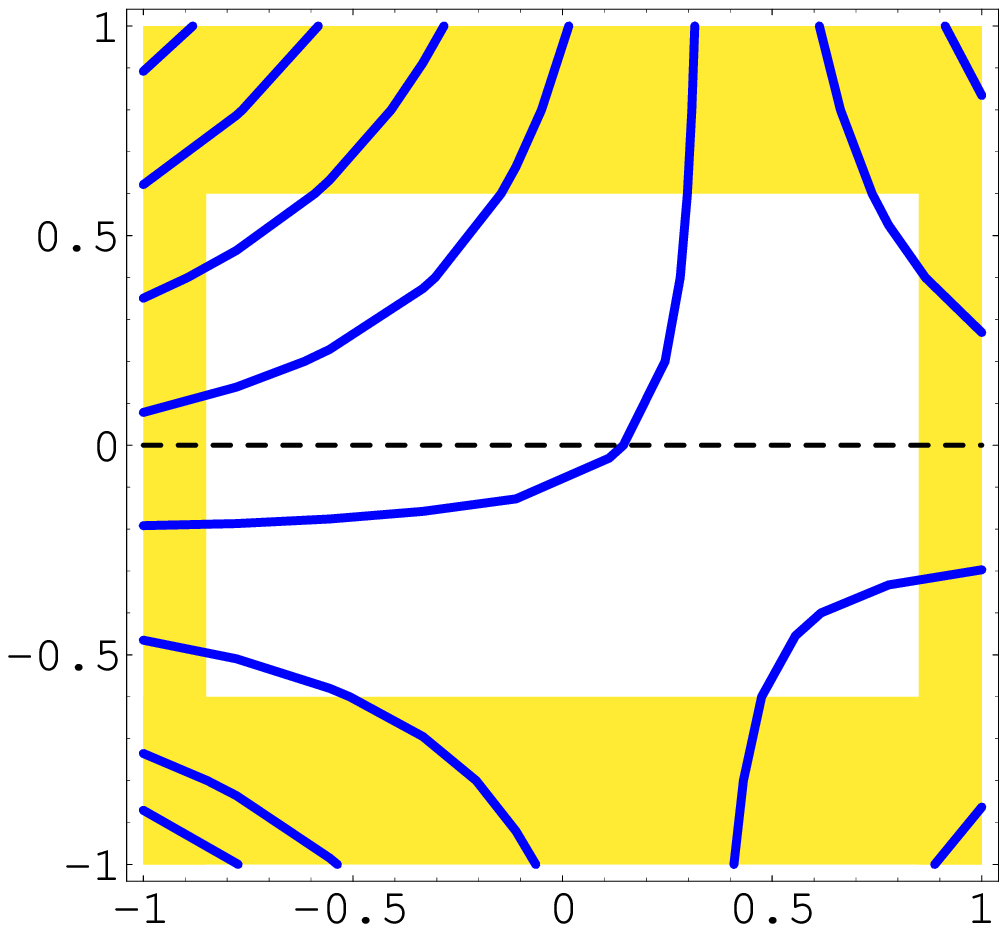}}
\put(3.3,5){a)}
\put(6.2,-1.2){ \small $ P_-^3$}
\put(-.3,4.6){\small $ P_+^3$}
\put(1.1,-.3){\tiny 5}
\put(1.3,0){\tiny 10}
\put(1.8,.7){\tiny 20}
\put(4.6,1){\tiny 30}
\put(5.5,-.3){\tiny 40}
\put(4.1,2.1){\tiny 30}
\put(2.7,2.75){\tiny 40}
\put(2,3.3){\tiny 50}
\put(1.5,3.7){\tiny 60}
\put(1,4.05){\tiny 70}
\put(4.9,3.3){\tiny 20}
\put(5.5,4){\tiny 10}
\end{picture}\par\vspace{1cm}
\end{minipage}\hfill\hspace{.2cm}
\begin{minipage}{7cm}
\begin{picture}(7,7)
\put(0,0){\includegraphics{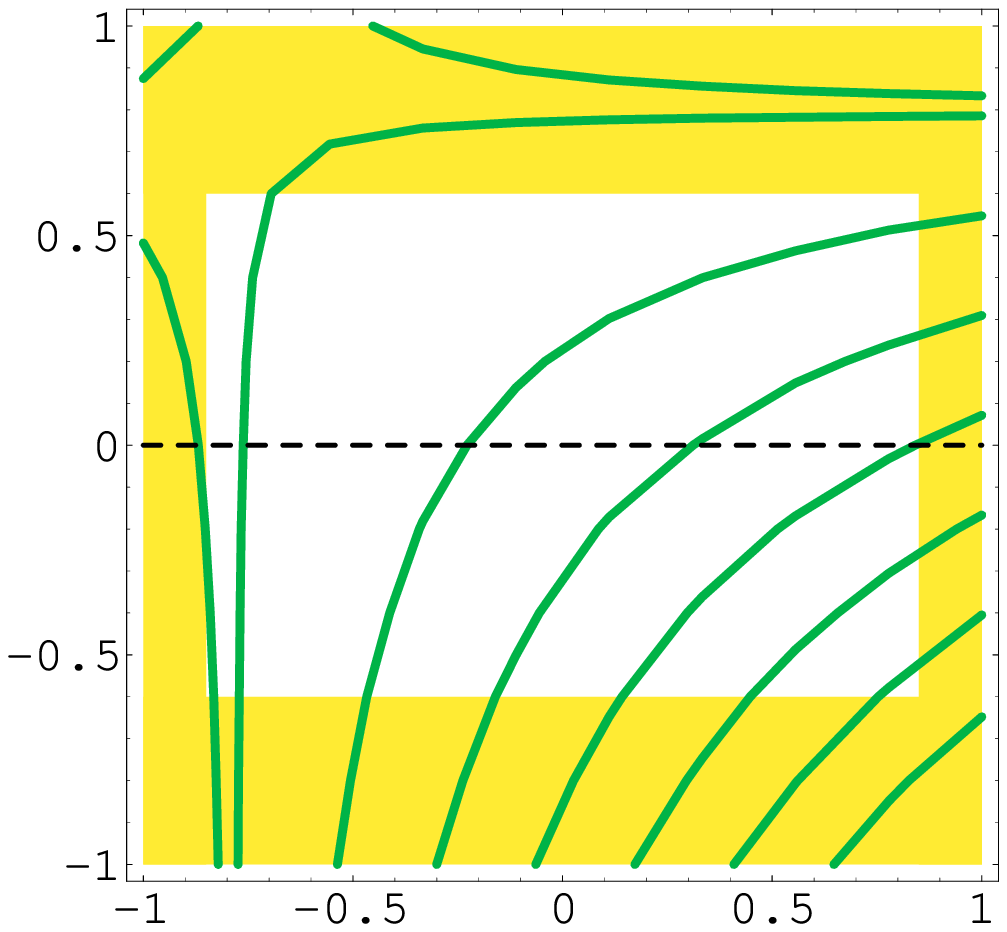}}
\put(3.3,5){b)}
\put(6.2,-1.2){\small $ P_-^3$}
\put(-.3,4.6){\small $ P_+^3$}
\put(1.1,4.1){\tiny 5}
\put(1,3.2){\tiny 4}
\put(2.1,4.2){\tiny 4}
\put(1.55,3.65){\tiny 5}
\put(5.25,0){\tiny 35}
\put(4.9,.3){\tiny 30}
\put(4.5,.7){\tiny 25}
\put(4,1.2){\tiny 20}
\put(3.55,1.7){\tiny 15}
\put(2.7,2.3){\tiny 10}
\end{picture}\par
\hspace{-.5cm}
\end{minipage}
\hspace*{-.8cm}
\begin{minipage}{7cm}
\vspace{-1cm}
\begin{center}
\begin{picture}(7,7)
\put(0,0){\includegraphics{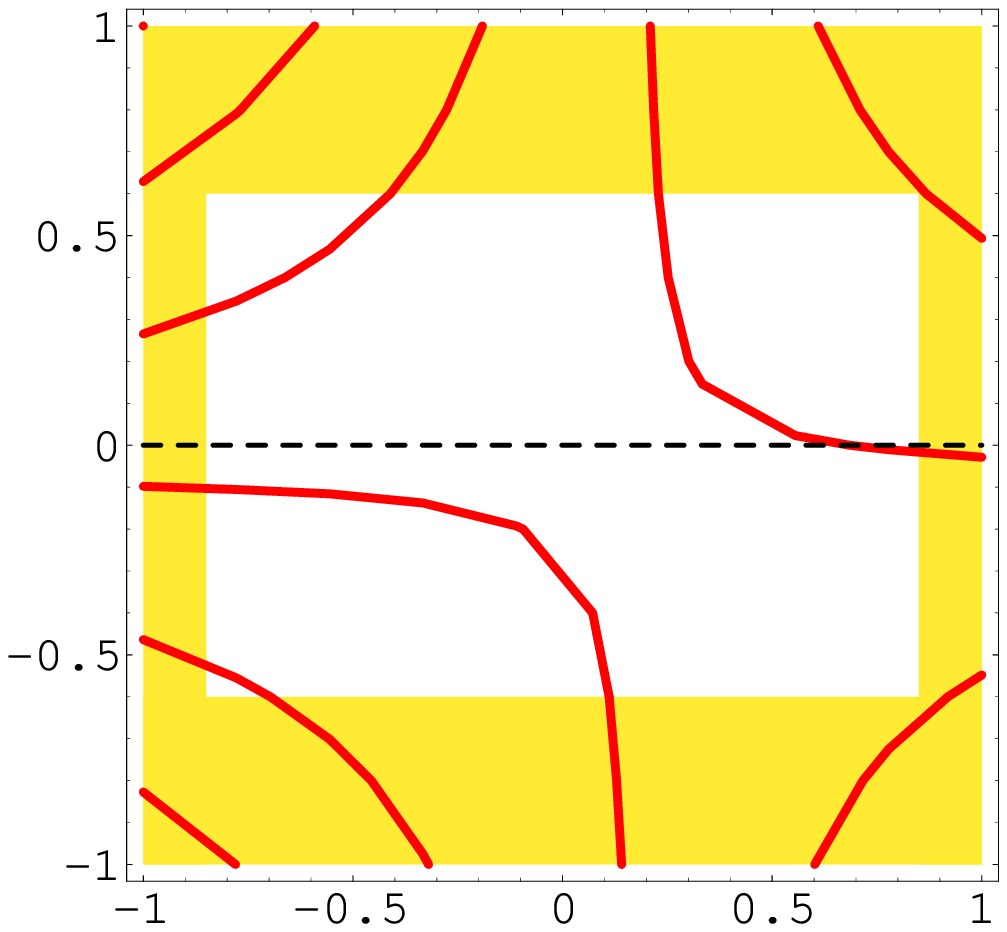}}
\put(3.3,5){c)}
\put(6.2,-1.2){\small $ P_-^3$}
\put(-.3,4.6){\small $ P_+^3$}
\put(5,3.5){\tiny $300$}
\put(3.7,2.1){\tiny 500}
\put(4.9,0.1){\tiny 700}
\put(2.3,3.1){\tiny 700}
\put(1.4,3.6){\tiny 900}
\put(1.2,-.3){\tiny 100}
\put(3.1,1.6){\tiny 500}
\put(1.8,.5){\tiny 300}
\end{picture}\par\vspace{.6cm}
\end{center}
\end{minipage}\hfill\hspace{.2cm}
\begin{minipage}{7cm}
\vspace{1.2cm}
{\parbox{6.5cm}{\small Figure 1:
Contour lines of cross sections 
$\sigma(e^+ e^-\to\tilde{\chi}^0_1\tilde{\chi}^0_2)$/fb 
at $\sqrt{s}=(m_{\tilde{\chi}^0_1}+m_{\tilde{\chi}^0_2})+30$~GeV
in a) scenario~A1 with $m_{\tilde{e}_L}=176$~GeV, $m_{\tilde{e}_R}=132$~GeV, 
in b) scenario~A2 with $m_{\tilde{e}_L}=500$~GeV, $m_{\tilde{e}_R}=132$~GeV, 
and in c) scenario~B with $m_{\tilde{e}_L}=176$~GeV, 
$m_{\tilde{e}_R}=132$~GeV. The longitudinal  
beam polarization for electrons (positrons) is denoted by $P_-^3$ ($P_+^3$).
The shaded region is for $|P_{-}^3|>85\%$, $|P_+^3|>60\%$
(dashed--line if only  electron beam polarized). }}
\end{minipage}
\end{figure}

\begin{figure}[t]
\hspace{-.8cm}
\begin{minipage}{7cm}
\begin{picture}(7,7)
\put(0,0){\includegraphics{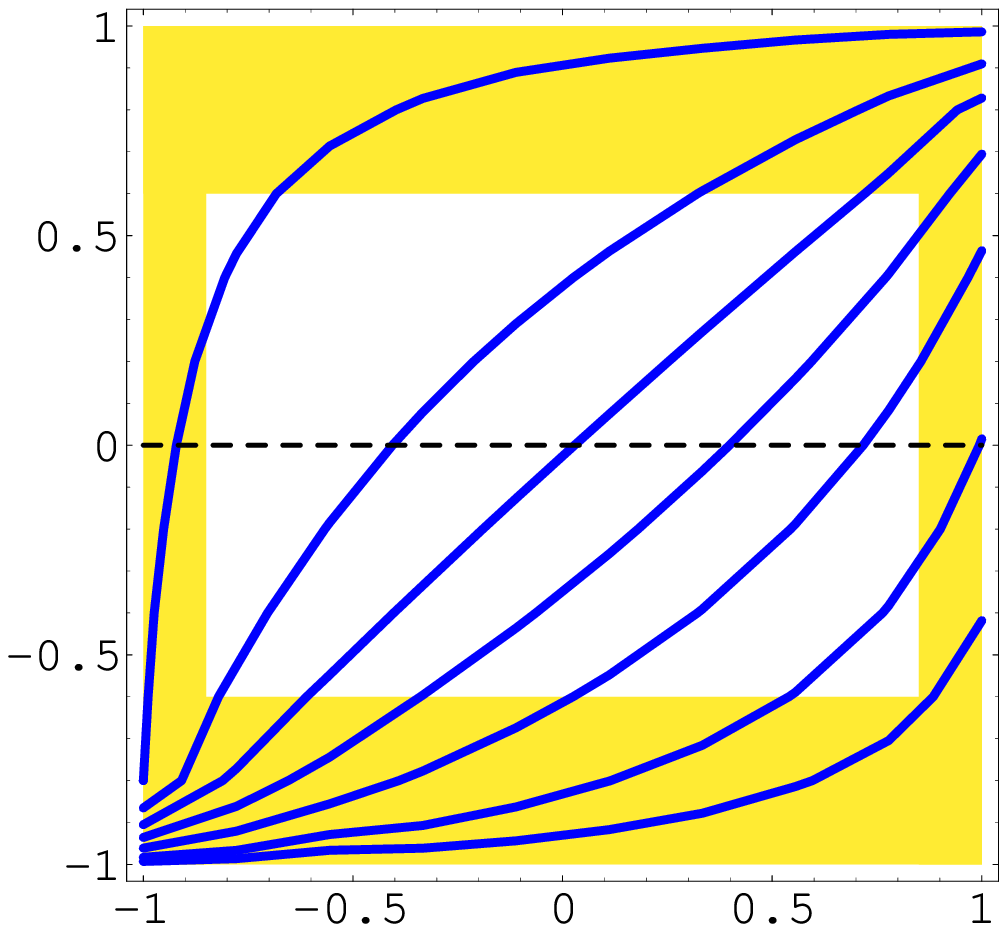}}
\put(3.3,5){a)}
\put(6.2,-1.2){\small $ P_-^3$}
\put(-.3,4.6){\small $ P_+^3$}
\put(3.8,1.6){\tiny 0}
\put(4.3,.3){\tiny 8}
\put(3.8,.8){\tiny 4}
\put(4.7,0){\tiny 10}
\put(3.2,2.1){\tiny $-4$}
\put(2.6,2.7){\tiny $-8$}
\put(1.2,3.8){\tiny $-12$}
\end{picture}\par\vspace{1cm}
\end{minipage}\hfill\hspace{.2cm}
\begin{minipage}{7cm}
\begin{picture}(7,7)
\put(0,0){\includegraphics{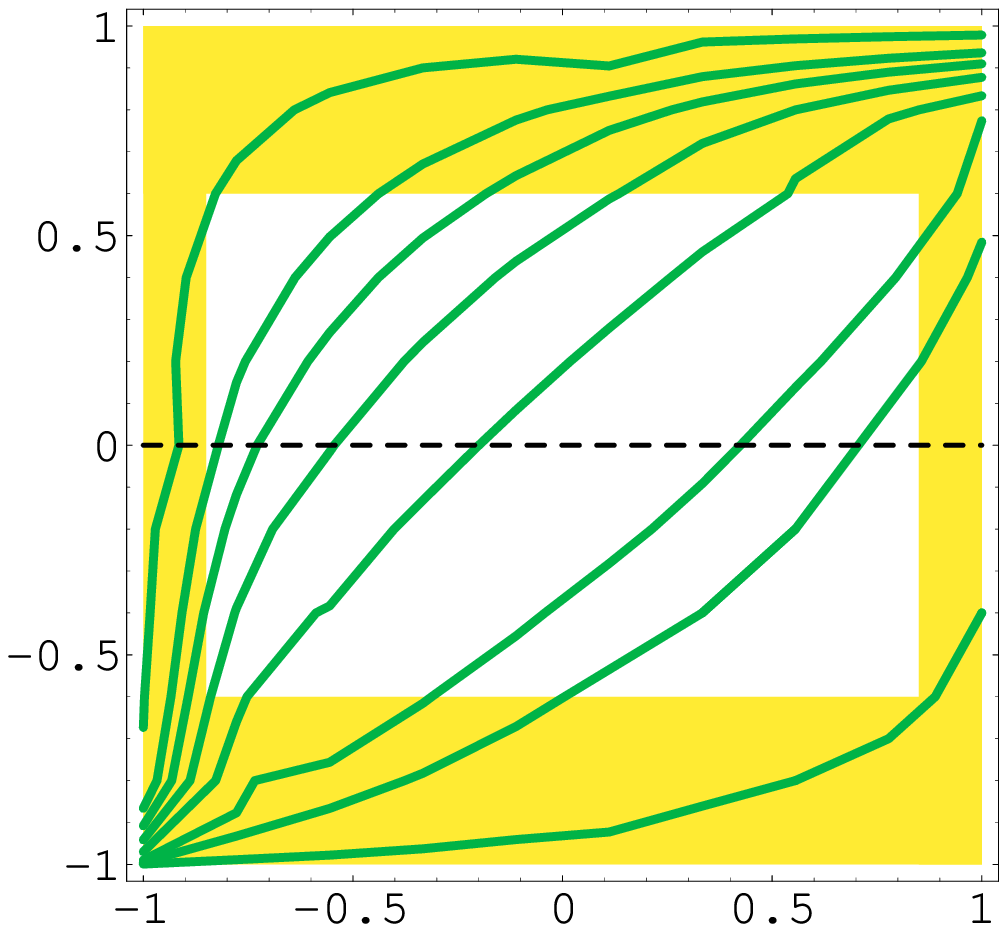}}
\put(3.3,5){b)}
\put(6.2,-1.2){\small $ P_-^3$}
\put(-.3,4.6){\small $ P_+^3$}
\put(4.9,.1){\tiny $16$}
\put(3.7,.8){\tiny $15$}
\put(3.3,1.2){\tiny $14$}
\put(2.7,2.1){\tiny 10}
\put(2.3,2.5){\tiny 5}
\put(2,2.9){\tiny 0}
\put(1.6,3.25){\tiny $-5$}
\put(1,3.85){\tiny $-15$}
\end{picture}\par
\hspace{-.5cm}
\end{minipage}
\hspace*{-.8cm}
\begin{minipage}{7cm}
\vspace{-1cm}
\begin{center}
\begin{picture}(7,7)
\put(0,0){\includegraphics{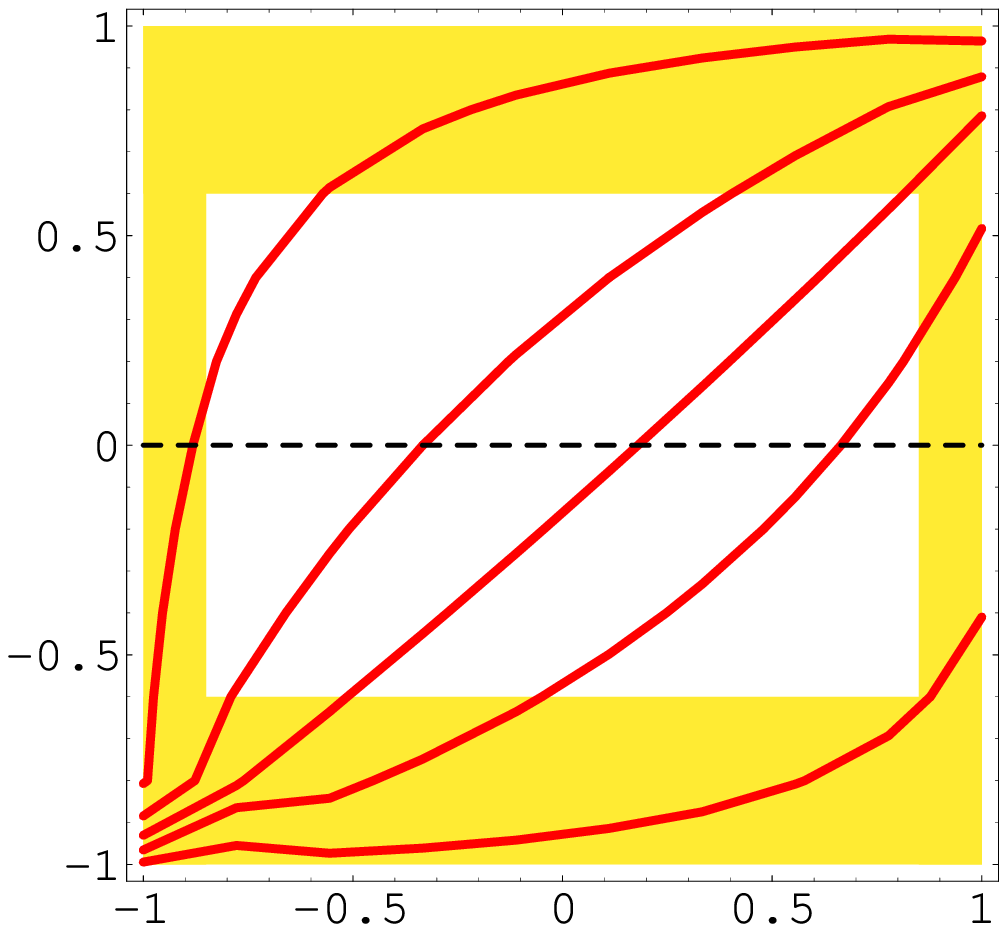}}
\put(3.3,5){c)}
\put(6.2,-1.2){\small $ P_-^3$}
\put(-.3,4.6){\small $ P_+^3$}
\put(4.7,.1){\tiny $-4$}
\put(3.1,1.5){\tiny 0}
\put(3.6,.9){\tiny $-2$}
\put(2.5,2.1){\tiny 2}
\put(1.5,3.2){\tiny 4}
\end{picture}\par\vspace{.6cm}
\end{center}
\end{minipage}\hfill\hspace{.2cm}
\begin{minipage}{7cm}
\vspace{1.4cm}
{\parbox{6.5cm}{\small Figure 2:
Contour lines of the forward--backward
asymmetry of the decay electron 
$A_{FB}$/\% of
$e^+ e^-\to\tilde{\chi}^0_1\tilde{\chi}^0_2, \tilde{\chi}^0_2\to 
\tilde{\chi}^0_1 e^+ e^-$
at $\sqrt{s}=(m_{\tilde{\chi}^0_1}+m_{\tilde{\chi}^0_2})+30$~GeV
in a) scenario~A1 ($m_{\tilde{e}_L}=176$~GeV, $m_{\tilde{e}_R}=132$~GeV), 
in b) scenario~A2 ($m_{\tilde{e}_L}=500$~GeV, $m_{\tilde{e}_R}=132$~GeV),
and in c)
scenario~B ($m_{\tilde{e}_L}=176$~GeV, $m_{\tilde{e}_L}=176$~GeV).
The longitudinal 
beam polarization for electrons (positrons) is denoted by $P_-^3$ ($P_+^3$).
The shaded region is for $|P_{-}^3|>85\%$, $|P_+^3|>60\%$
(dashed--line if only electron beam polarized).\la{fig_4}} }
\end{minipage}
\end{figure}

\begin{figure}[t]
\hspace{-.9cm}
\begin{minipage}{7cm}
\begin{picture}(7,5)
\put(0,0){\includegraphics{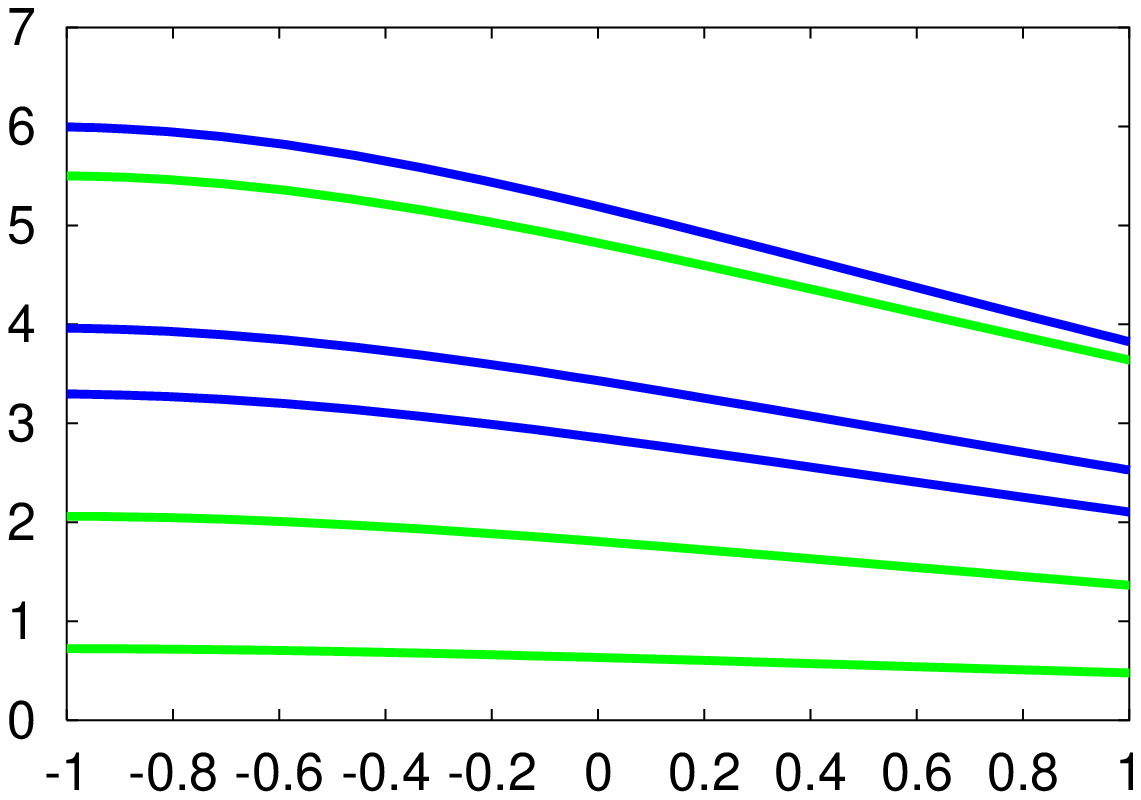}}
\put(3.6,5.4){a)}
\put(5.8,-.2){\small $ \cos\Theta_{+-}$}
\put(-.4,5.4){\small $ \frac{d\sigma_e}{d\cos\Theta_{+-}}$/fb}
\put(1,3.1){\tiny A1, $(+-)$}
\put(1,3.6){\tiny A2, $(+-)$}
\put(1,4.4){\tiny A1, $(-+)$}
\put(1,2.7){\tiny A1, $(00)$}
\put(1,2){\tiny A2, $(00)$}
\put(1,1.2){\tiny A2, $(-+)$}
\end{picture}\par
\end{minipage}\hfill\hspace{.2cm}
\begin{minipage}{7cm}
\begin{picture}(7,5)
\put(0,0){\includegraphics{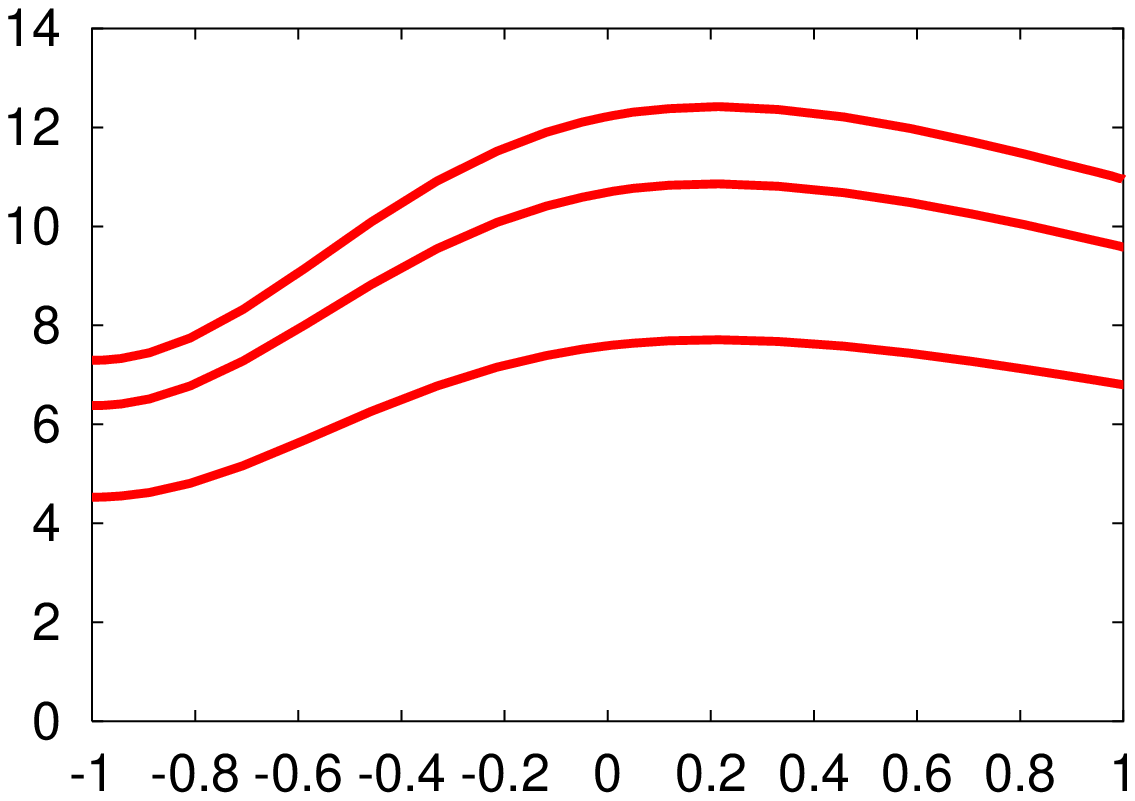}}
\put(3.6,5.4){b)}
\put(5.8,-.2){\small $ \cos\Theta_{+-}$}
\put(-.4,5.4){\small $ \frac{d\sigma_e}{d\cos\Theta_{+-}}$/fb}
\put(.8,1.7){\tiny B, $(00)$}
\put(2.4,3){\tiny B, $(+-)$}
\put(.8,3.5){\tiny B, $(-+)$}
\end{picture}\par
\end{minipage}
\addtocounter{figure}{2}
\caption{Opening angle distribution of the decay 
leptons from $e^+ e^-\to \tilde{\chi}^0_1 \tilde{\chi}^0_2$, 
$\tilde{\chi}^0_2\to e^+ e^- \tilde{\chi}^0_1$ in a)
scenario~A1 ($m_{\tilde{e}_R}=132$~GeV, $m_{\tilde{e}_L}=176$~GeV),  
A2 ($m_{\tilde{e}_R}=132$~GeV, $m_{\tilde{e}_L}=500$~GeV)
and in b) scenario~B ($m_{\tilde{e}_R}=132$~GeV, 
$m_{\tilde{e}_L}=176$~GeV) 
at $\sqrt{s}=(m_{\tilde{\chi}^0_1}+m_{\tilde{\chi}^0_2})+30$~GeV  
for unpolarized beams (00), for $P_{-}=-85$\%, $P_{+}=+60$\% $(-+)$ and
for $P_{-}=+85$\%, $P_{+}=-60$\% $(+-)$, respectively.\la{fig_5}}
\end{figure}

\subsection{Dependence on $M_1$}\la{sec:34}
So far we have used the GUT relation (\re{eq_m1})
for the gaugino masses.
In the following we will be more general and not use
this relation \ci{Choi,Feng,Snowmass,Kneur}. We will discuss the 
$M_1$
dependence of the cross section and the forward--backward asymmetry
of the decay electron \ci{Gudi_neut}.
All other parameters are chosen as in scenario~A1 except the mass 
of $\tilde{e}_R$, $m_{\tilde{e}_R}=161$~GeV.
The neutralino masses as well as the $Z^0 \tilde{\chi}^0_i \tilde{\chi}^0_j$
couplings 
$O_{ij}^{''L}$ and the $\tilde{\chi}^0_i \tilde{\ell} \ell$
couplings $f^{L,R}_{\ell i}$ 
depend on $M_1$.

In Fig.~\re{fig_6} we show the neutralino masses as function of $M_1$.
The grey areas are excluded by 
the constraints $m_{\tilde{\chi}^0_1}<m_{\tilde{\chi}^{\pm}_1}$,
$m_{\tilde{\chi}^0_1}>35$~GeV. We see that in the interval 
between $-$130~GeV$<M_1<M_2$  $m_{\tilde{\chi}^0_1}$ depends very 
strongly on $M_1$, whereas all other neutralino masses are nearly independent
of $M_1$. On the 
other hand, in the region $M_2\le M_1 \le |\mu|$
only $m_{\tilde{\chi}^0_2}$ depends strongly on $M_1$.

In the formulae for the cross section
$\sigma(e^+e^-\to \tilde{\chi}^0_1 \tilde{\chi}^0_2)$ and for the decay
$\tilde{\chi}^0_2\to \tilde{\chi}^0_1 \ell^+ \ell^-$ the products of 
couplings $f^L_{\ell 1}f^L_{\ell 2}$ and $f^R_{\ell 1} f^R_{\ell 2}$ 
enter. We therefore show for these products 
the dependence on $M_1$ in Fig.~\re{fig_7}.
We do not consider values $|M_1|>160$~GeV, where 
$m_{\tilde{\chi}^0_2}>m_{\tilde{e}_R}$, because then the two--body decay
$\tilde{\chi}^0_2\to \tilde{e}_R+e$ would be possible.
One observes a strong variation of $f^R_{\ell 1} f^R_{\ell 2}$ and 
$f^L_{\ell 1} f^L_{\ell 2}$ for 
$M_1>80$~GeV. In particular, $f^R_{\ell 1} f^R_{\ell 2}$ 
has a positive maximum at 140~GeV, 
whereas $f^L_{\ell 1} f^L_{\ell 2}$ is zero at $M_1=120$~GeV and reaches large 
negative values for $M_1\ge 160$~GeV. We therefore have the following regions:
$f^L_{\ell 1} f^L_{\ell 2}>f^R_{\ell 1} f^R_{\ell 2}>0$ for 
$-200$~GeV$<M_1<80$~GeV, 
$|f^R_{\ell 1} f^R_{\ell 2}|>f^L_{\ell 1} f^L_{\ell 2}$ for
110~GeV$<M_1<140$~GeV, and $|f^L_{\ell 1} f^L_{\ell 2}|>f^R_{\ell 1} 
f^R_{\ell 2}$ for $M_1>150$~GeV.
The coupling $O_{12}^{''L}$ is small in this gaugino--like scenario.

\bsl
Fig.~\re{fig_8}a exhibits the $M_1$ dependence of $\sigma(e^+e^-\to 
\tilde{\chi}^0_1\tilde{\chi}^0_2)\times BR(\tilde{\chi}^0_2\to e^+ e^- 
\tilde{\chi}^0_1)$ at $\sqrt{s}=(m_{\tilde{\chi}^0_1}+m_{\tilde{\chi}^0_2})+
30$~GeV in the region 40 GeV$<M_1<$160~GeV
for various beam polarizations.
Since the masses of $\tilde{e}_L$ and $\tilde{e}_R$ are in this case 
comparable, the curves reflect the behaviour of $f^L_{\ell 1}f^L_{\ell 2}$ 
and $f^R_{\ell 1} f^R_{\ell 2}$ 
of Fig.~\re{fig_7}. 
A left (right) beam polarization of the electron (positron) selects
the $\tilde{e}_L$ exchange, while the maximum in the curve with  a right
(left) electron (positron) polarization is due the maximum of 
$f^R_{\ell 1} f^R_{\ell 2}$ 
in Fig.~\re{fig_7}.
\esl

Fig.~\re{fig_8}b shows the analogous curves for a heavy $\tilde{e}_L$ 
($m_{\tilde{e}_L}=500$~GeV)
and all other parameters as in Fig.~\re{fig_8}a. 
One clearly sees that the 
$\tilde{e}_L$ exchange is strongly suppressed, and one obtains higher cross 
sections for right polarized $e^-$ beams.

We have also studied the $M_1$ 
dependence of the forward--backward 
asymmetry $A_{FB}$ of the decay electron, 
eq.~(\re{eq_30}). It is shown in Fig.~\re{fig_9}a for 
$m_{\tilde{e}_L}=176$~GeV, and in Fig.~\re{fig_9}b
for $m_{\tilde{e}_L}=500$~GeV. 
One notices a strong variation with $M_1$ and a strong dependence on the beam 
polarizations. Comparing Fig.~\re{fig_9}a and Fig.~\re{fig_9}b,
one observes a very pronounced difference of the forward--backward asymmetry 
of the decay electron in the region 
40~GeV$<M_1<$100~GeV. This is 
due to the suppression of $\tilde{e}_L$ exchange in 
Fig.~\re{fig_9}b. 
The beam polarizations enhance the effect considerably.
The peak at $M_1\approx 120$~GeV is again due to the maximum of 
$f^R_{\ell 1} f^R_{\ell 2}$.

\begin{figure}
\begin{center}
\begin{picture}(11,9)
\put(-.5,-1.3){\includegraphics{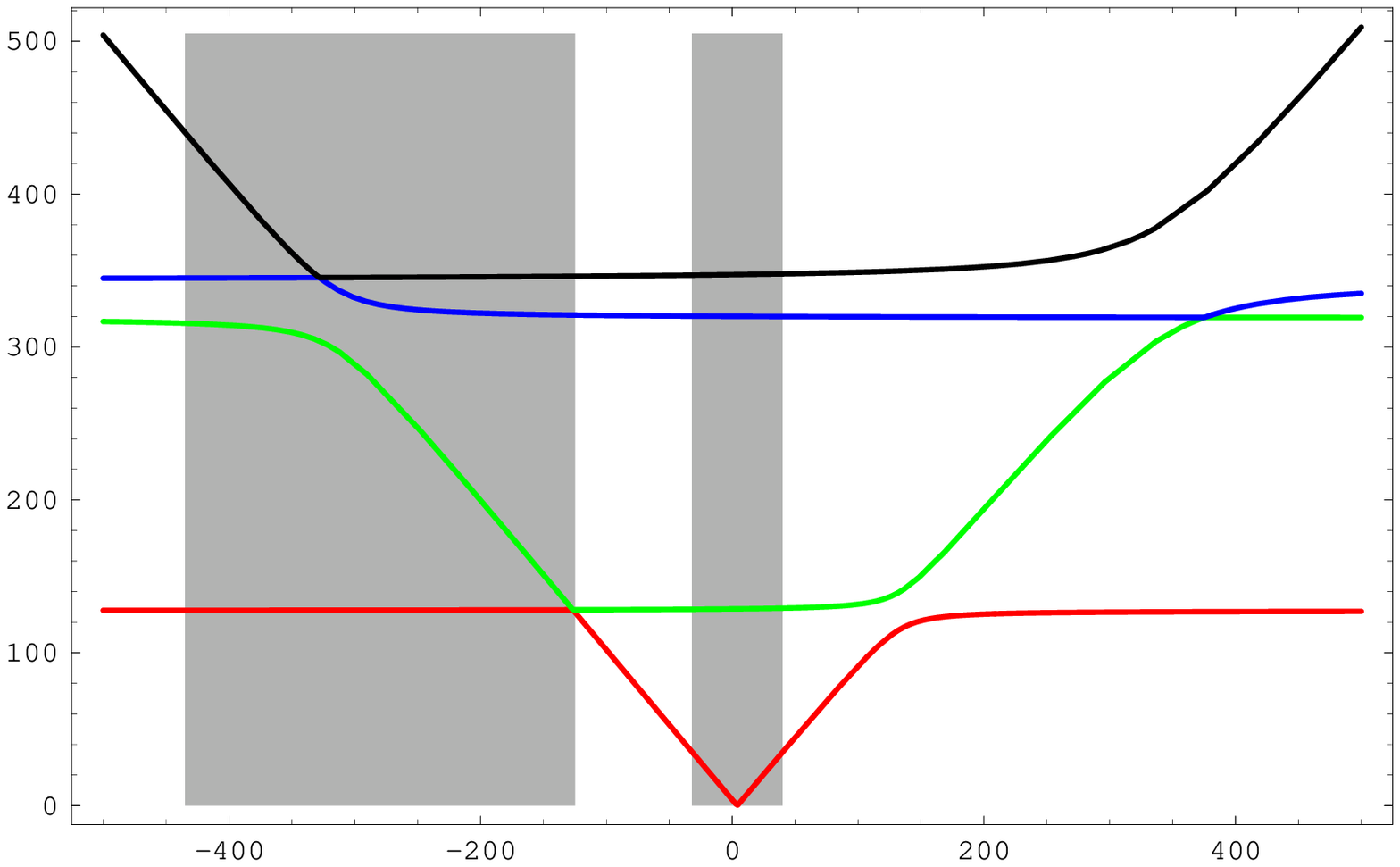}}
\put(-1.2,8.6){\small [GeV]}
\put(10,1){$M_1$\small /GeV}
\put(6.75,1.5){\tiny $\uparrow$}
\put(6.5,1.1){\small GUT: 78.7}
\put(-1,3.5){\small $m_{\tilde{\chi}^0_1}$}
\put(-1,5.8){\small $m_{\tilde{\chi}^0_2}$}
\put(-1,6.2){\small $m_{\tilde{\chi}^0_3}$}
\put(-1,8.2){\small $m_{\tilde{\chi}^0_4}$}
\end{picture}\par\vspace{-1.3cm}
\caption{Neutralino mass spectrum as a function of  
$M_1$ in the range $-500$~GeV$<M_1<500$~GeV. 
The grey areas are excluded by 
the constraints $m_{\tilde{\chi}^0_1}<m_{\tilde{\chi}^{\pm}_1}$,
$m_{\tilde{\chi}^0_1}>35$~GeV.
All other MSSM parameters as in scenario~A1;
GUT relation (\re{eq_m1}) gives $M_1=78.7$~GeV.\la{fig_6}}
\end{center}
\end{figure}
\begin{figure}
\begin{picture}(11,8)
\put(10,1.1){$M_1$\small /GeV}
\put(-.5,-1.3){\includegraphics{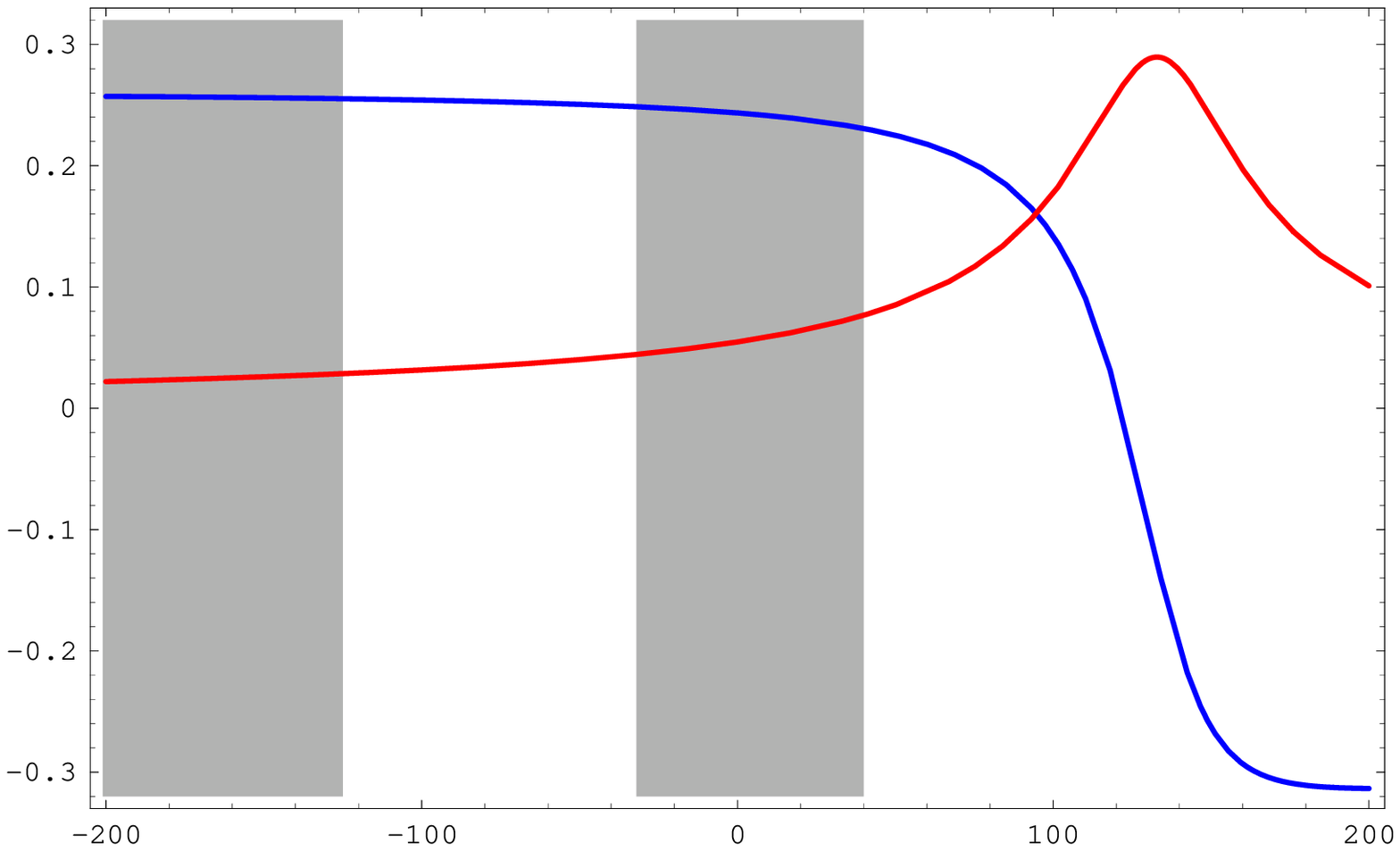}}
\put(-.4,7.7){\small $ f^L_{\ell 1} f^L_{\ell 2}$}
\put(-.4,5.2){\small $ f^R_{\ell 1} f^R_{\ell 2}$}
\end{picture}\par\vspace{-1.4cm}
\caption{Slepton couplings 
$f^L_{\ell 1} f^L_{\ell 2}$ and $f^R_{\ell 1} f^R_{\ell 2}$ as a function of  
$M_1$ in the range $-200$~GeV$<M_1<200$~GeV.
All other MSSM parameters as in scenario~A1;
GUT relation (\re{eq_m1}) gives $M_1=78.7$~GeV.\la{fig_7}}
\end{figure}

\begin{figure}
\hspace{-.9cm}
\begin{minipage}{7cm}
\begin{picture}(7,5)
\put(0,0){\includegraphics{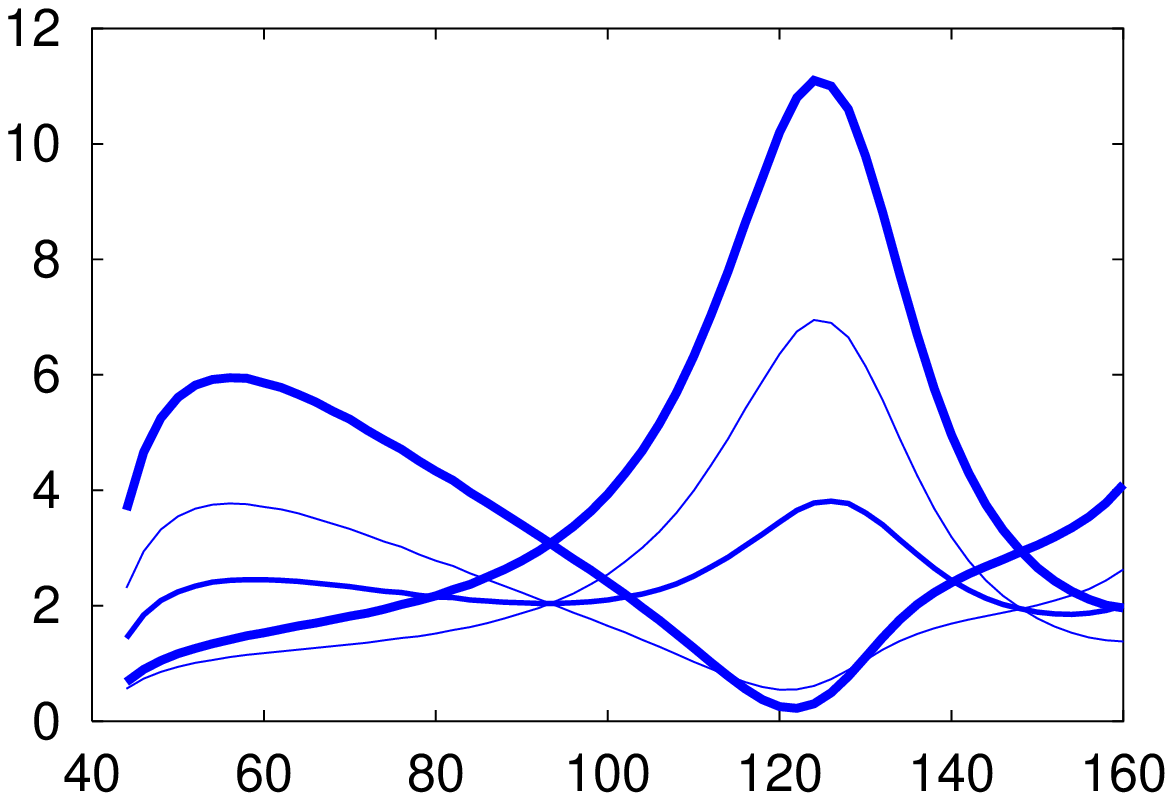}}
\put(3.5,5.2){a)}
\put(5.8,-.3){ \small $M_1${\small /GeV}}
\put(-.4,5.2){\small $ \sigma_e${\small /fb}}
\put(4.7,2.1){\small $(00)$}
\put(4.6,3.3){\small $(+0)$}
\put(5.5,4){\small $(+-)$}
\put(1.1,2.1){\small $(-0)$}
\put(1.1,2.9){\small $(-+)$}
\end{picture}\par
\end{minipage}\hfill\hspace{.2cm}
\begin{minipage}{7cm}
\begin{picture}(7,5)
\put(0,0){\includegraphics{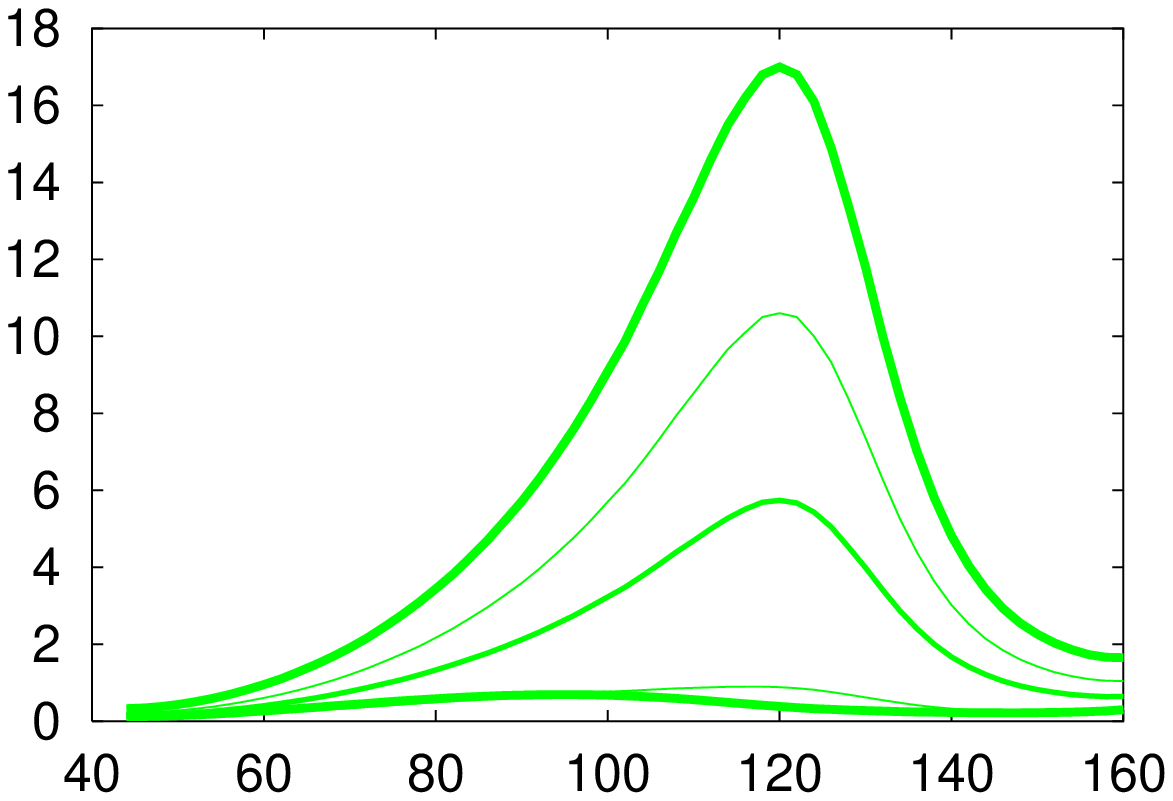}}
\put(3.5,5.2){b)}
\put(5.8,-.3){\small $ M_1${\small /GeV}}
\put(-.4,5.2){\small $ \sigma_e$ {\small /fb}}
\put(3.5,1){\small $(-+)\approx (-0)$}
\put(4.4,2.1){\small $(00)$}
\put(4.4,3.3){\small $(+0)$}
\put(5.3,4.2){\small $(+-)$}
\end{picture}\par
\end{minipage}
\caption{Cross sections 
$\sigma_e=\sigma(e^+e^-\to\tilde{\chi}^0_1\tilde{\chi}^0_2)\times 
BR(\tilde{\chi}^0_2\to \tilde{\chi}^0_1 e^+ e^-)$ at $\sqrt{s}=
m_{\tilde{\chi}^0_1}+m_{\tilde{\chi}^0_2}+30$~GeV as function of gaugino 
parameter $M_1$ for unpolarized beams (00), 
for only electron beam polarized 
$(-0)$, $(+0)$ with $P_-^3=\pm 85\%$ and for both beams 
polarized $(-+)$, $(+-)$ 
with $P_-=\mp 85\%$, $P_+=\pm 60\%$. The slepton masses are  a) 
$m_{\tilde{e}_L}=176$~GeV, $m_{\tilde{e}_R}=161$~GeV, and b) 
$m_{\tilde{e}_L}=500$~GeV, $m_{\tilde{e}_R}=161$~GeV; the
other SUSY parameters as in scenario~A1.\la{fig_8}}
\end{figure}

\begin{figure}[t]
\hspace{-.9cm}
\begin{minipage}{7cm}
\begin{picture}(7,5)
\put(-.2,0){\includegraphics{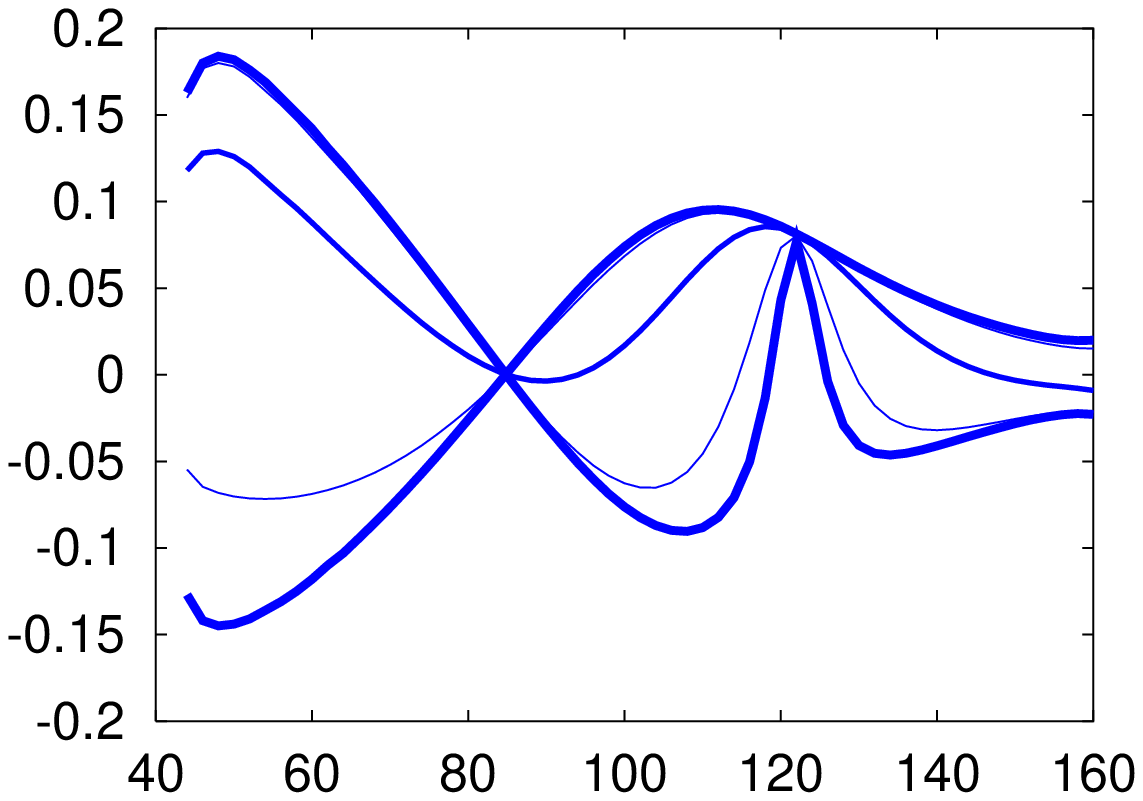}}
\put(3.5,5.2){a)}
\put(5.5,-.3){ \small $ M_1${\small /GeV}}
\put(-.2,5.2){\small $ A_{FB}$}
\put(1.9,4.4){\small $(-+)\approx (-0)$}
\put(4,1.4){\small $(-+)$}
\put(3.6,2.4){\small $(-0)$}
\put(4.5,4){\small $(+-)\approx (+0)$}
\put(1.1,3.5){\small $(00)$}
\put(1.2,2.2){\small $(+0)$}
\put(1.9,1){\small $(+-)$}
\end{picture}\par
\end{minipage}\hfill\hspace{.2cm}
\begin{minipage}{7cm}
\begin{picture}(7,5)
\put(-.2,0){\includegraphics{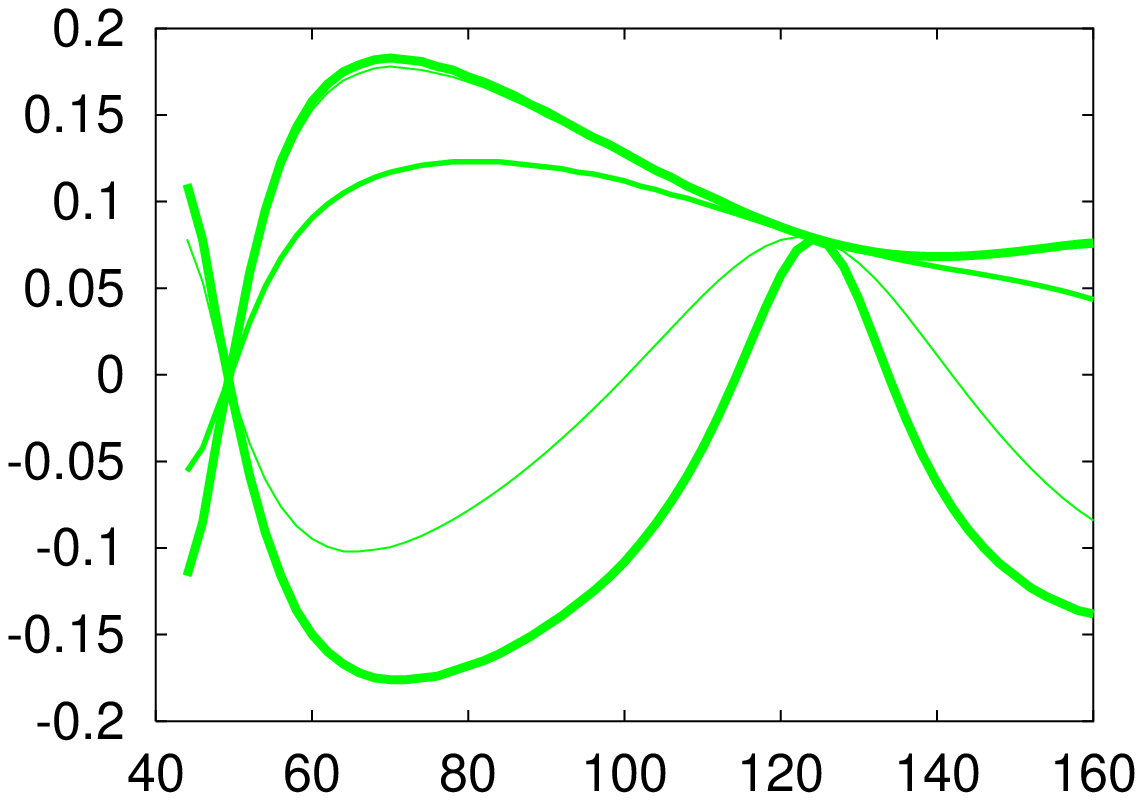}}
\put(3.5,5.2){b)}
\put(5.5,-.3){\small $ M_1/${\small ~GeV}}
\put(-.2,5.2){\small $ A_{FB}$}
\put(3.5,4.4){\small $(+-)\approx (+0)$}
\put(2.1,3.4){\small $(00)$}
\put(2,2){\small $(-0)$}
\put(2.1,1.1){\small $(-+)$}
\end{picture}\par\vspace{.1cm}
\end{minipage}
\caption{Forward--backward asymmetry of decay electron 
$A_{FB}$ of $e^+ e^-\to \tilde{\chi}^0_1 \tilde{\chi}^0_2$, 
$\tilde{\chi}^0_2\to \tilde{\chi}^0_1 e^+ e^- $
at $\sqrt{s}=m_{\tilde{\chi}^0_1}+m_{\tilde{\chi}^0_2}+30$~GeV as 
function of gaugino 
parameter $M_1$ for unpolarized beams (00), 
for only electron beam polarized 
$(-0)$, $(+0)$ 
with $P_-^3=\pm 85\%$ and for both beams polarized $(-+)$, 
$(+-)$ 
with $P_-=\mp 85\%$, $P_+=\pm 60\%$. Slepton masses are a)  
$m_{\tilde{e}_L}=176$~GeV, $m_{\tilde{e}_R}=161$~GeV, and b) 
$m_{\tilde{e}_L}=500$~GeV, $m_{\tilde{e}_R}=161$~GeV; 
the other SUSY parameters as in scenario~A1. \la{fig_9}}
\end{figure}

\section{Conclusions}
The objective of this paper has been twofold. Firstly, we have studied 
the advantage of having both the $e^-$ and the $e^+$ beam polarized. 
If the polarizations of $e^-$ and $e^+$ are varied,
the relative size of the cross sections depends significantly 
on the mixing character of the neutralinos and on 
the masses of $\tilde{e}_L$ and $\tilde{e}_R$.
By an appropriate choice of the polarizations
one can obtain up to three times larger cross sections than in the unpolarized
case. 
 Secondly, 
by taking into account the full spin correlations between production 
and decay, we have studied the 
angular distribution, as well as the forward--backward asymmetry of 
the decay electron 
$e^+ e^-\to \tilde{\chi}^0_1 \tilde{\chi}^0_2$, 
$\tilde{\chi}^0_2\to \tilde{\chi}^0_1 e^+ e^-$. Measuring this asymmetry 
for various beam polarizations strongly constrains the masses of 
$\tilde{e}_L$ and $\tilde{e}_R$ and the mixing properties of 
the neutralinos. 
We have also
studied the dependence on the gaugino mass parameter $M_1$. For a 
determination of $M_1$ the use of polarized $e^+$ and 
$e^-$ beams would be very 
helpful. Due to the Majorana character of the neutralinos the opening angle 
distribution between the decay leptons is independent of spin correlations.
It is very sensitive to the mixing character of the neutralinos, whereas its 
shape is only weakly dependent on the selectron masses.
\vspace{-1mm}
\section*{Acknowledgments}
\vspace{-1mm}
We are grateful to W.~Porod and S.~Hesselbach 
for providing the computer programs for neutralino widths.
G.M.-P. was supported by {\it
  Friedrich-Ebert-Stiftung}. This work was also supported by 
the German Federal Ministry for
Research and Technology (BMBF) under contract number
05 7WZ91P (0), by the Deutsche Forschungsgemeinschaft under
contract Fr 1064/4-1, and the `Fonds zur
F\"orderung der wissenschaftlichen Forschung' of Austria, Project
No. P13139-PHY.

\end{document}